\documentclass[twoside]{ilcws07}
\usepackage[latin1]{inputenc}
\usepackage[dvips]{graphicx,epsfig,color}
\usepackage{wrapfig,rotating}
\usepackage{amssymb,amsmath,array}

\input{paperdef}

\begin{document}

\thispagestyle{empty}
\setcounter{page}{0}
\def\thefootnote{\fnsymbol{footnote}}

\begin{flushright}
DCPT/07/112 \\
IPPP/07/56 \\
MPP--2007--123 
\end{flushright}

\vspace{1cm}

\begin{center}

{\large\sc {\bf Consistent Treatment of Imaginary Contributions\\[.5em]
to Higgs-Boson Masses in the MSSM}}
\footnote{talk given by S.~Heinemeyer at the {\em LCWS07}, 
May 2007, DESY, Hamburg, Germany}

\vspace{1cm}

{\sc 

T.~Hahn$^{1\,}$%
\footnote{
email: hahn@feynarts.de}
, S.~Heinemeyer$^{2\,}$%
\footnote{
email: Sven.Heinemeyer@cern.ch}
, W.~Hollik$^{1\,}$%
\footnote{
email: hollik@mppmu.mpg.de}
, H.~Rzehak$^{3\,}$%
\footnote{
email: Heidi.Rzehak@psi.ch}
~and G.~Weiglein$^{4\,}$%
\footnote{
email: Georg.Weiglein@durham.ac.uk}

}

\vspace*{1cm}

{\it
$^1$Max-Planck-Institut f\"ur Physik, 
F\"ohringer Ring 6, D--80805 Munich, Germany\\[.3em]
$^2$Instituto de Fisica de Cantabria (CSIC-UC), 
Santander,  Spain \\[.3em]
$^3$Paul Scherrer Institut, W\"urenlingen und Villigen, CH--5232
Villigen PSI, Switzerland\\[.3em]
$^4${IPPP, University of Durham, Durham DH1~3LE, UK}
}
\end{center}

\vspace*{0.2cm}

\BC {\bf Abstract} \EC
We show how the imaginary parts of the Higgs-Boson self-energies in
the MSSM are consistenly taken into account in the Higgs-Boson mass
determination. In a numerical example we find effects of $5 \gev$ in the 
mass difference of the two heavy neutral Higgs bosons.
The imaginary contributions have been included into the code {\tt FeynHiggs}.

\def\thefootnote{\arabic{footnote}}
\setcounter{footnote}{0}

\newpage


\graphicspath{{figs/}}

\title{
Consistent Treatment of Imaginary Contributions\\
to Higgs-Boson Masses in the MSSM}
\author{T.~Hahn$^1$, S.~Heinemeyer$^2$, W.~Hollik$^1$,
H.~Rzehak$^3$ and G.~Weiglein$^4$
\vspace{.3cm}\\
1- Max-Planck-Institut f\"ur Physik, 
F\"ohringer Ring 6, D--80805 Munich, Germany \\[.1cm]
2- Instituto de Fisica de Cantabria (CSIC-UC), Santander,  Spain \\[.1cm]
3- Paul Scherrer Institut, W\"urenlingen und Villigen, CH--5232
Villigen PSI, Switzerland\\[.1cm]
4- IPPP, University of Durham, Durham DH1~3LE, UK
}

\maketitle

\begin{abstract}
We show how the imaginary parts of the Higgs-Boson self-energies in
the MSSM are consistenly taken into account in the Higgs-Boson mass
determination. In a numerical example we find effects of $5 \gev$ in the 
mass difference of the two heavy neutral Higgs bosons.
The imaginary contributions have been included into the code {\tt FeynHiggs}.

\end{abstract}

\section{Introduction}

A striking prediction of models of 
supersymmetry (SUSY)~\cite{mssm} is a Higgs sector with at
least one relatively light Higgs boson. In the Minimal Supersymmetric
extension of the Standard Model (MSSM) two Higgs doublets are required,
resulting in five physical Higgs bosons: the light and heavy $\cp$-even $h$
and $H$, the $\cp$-odd $A$, and the charged Higgs bosons $H^\pm$.
The Higgs sector of the MSSM can be expressed at lowest
order in terms of $\MZ$, $\MA$ and $\tb \equiv v_2/v_1$, 
the ratio of the two vacuum expectation values. All other masses and
mixing angles can therefore be predicted. Higher-order contributions give 
large corrections to the tree-level relations.
The limits obtained from the Higgs search at LEP (the final LEP
results can be found in \citeres{LEPHiggsSM,LEPHiggsMSSM}), place
important restrictions on the parameter space of the MSSM.

For the MSSM with real parameters (rMSSM) the status of higher-order
corrections to the masses and mixing angles in the Higgs sector 
is quite advanced. The complete one-loop result within the rMSSM is
known~\cite{ERZ,mhiggsf1lA,mhiggsf1lB,mhiggsf1lC}.
The computation of the two-loop corrections
has meanwhile reached a stage where all the presumably dominant
contributions are available, see \citeres{mhiggsAEC,mhiggsEP5} and references
therein. Leading three-loop corrections have recently been obtained in
\citere{mhiggs3L}. 
The remaining theoretical uncertainty on the lightest $\cp$-even Higgs
boson mass has been estimated to be below 
$\sim 3 \gev$~\cite{mhiggsAEC,PomssmRep,mhiggsWN}. 
The public code \fh~\cite{mhiggsAEC,mhiggslong,feynhiggs,mhcMSSMlong}
is based on the results obtained in the Feynman-diagrammatic (FD)
approach~\cite{mhiggsAEC,mhiggslong,mhiggsletter,mhiggsEP4b}; it
includes all available corrections in the FD approach.
For the MSSM with complex parameters (cMSSM) the full one-loop result
in the FD approach has been obtained in \citere{mhcMSSMlong}, and the
corresponding leading \order{\alt\als} corrections can be found in
\citere{mhcMSSM2L}.


\section{Imaginary Contributions to Higgs-boson self-energies}

The propagator matrix of the neutral Higgs bosons $h, H, A$ can be
written as a $3 \times 3$ matrix, $\De_{hHA}(p^2)$.
The $3 \times 3$ propagator matrix
is related to the $3 \times 3$ matrix of the irreducible
vertex functions by
\begin{equation}
\De_{hHA}(p^2) = - \left(\hat{\Gamma}_{hHA}(p^2)\right)^{-1} ,
\label{eq:propagator}
\end{equation}
where 
\begin{align}
\label{eq:invprophiggs}
  \hat{\Gamma}_{hHA}(p^2) &= i \left[p^2 \unity - \matr{M}_{\mathrm{n}}(p^2)
                               \right], \\[.5em]
  \matr{M}_{\mathrm{n}}(p^2) &=
  \begin{pmatrix}
    \mh^2 - \ser{hh}(p^2) & - \ser{hH}(p^2) & - \ser{hA}(p^2) \\
    - \ser{hH}(p^2) & \mH^2 - \ser{HH}(p^2) & - \ser{HA}(p^2) \\
    - \ser{hA}(p^2) & - \ser{HA}(p^2) & \mA^2 - \ser{AA}(p^2)
  \end{pmatrix}. 
\label{eq:Mn}
\end{align}
The three complex poles ${\cal M}^2$ of $\De_{hHA}$, \refeq{eq:propagator},
are  determined as the solutions of
\begin{equation}
{\cal M}_i^2 - m_i^2 + \ser{ii}^{\rm eff}({\cal M}_i^2) = 0~,~i = h,H,A~.
\label{eq:massmaster}
\end{equation}
The effective self-energy reads (no summation over 
$i, j, k$)
\begin{align}
\ser{ii}^{\rm eff}(p^2) &= \ser{ii}(p^2) - i 
\frac{2 \hat{\Gamma}_{ij}(p^2) \hat{\Gamma}_{jk}(p^2) \hat{\Gamma}_{ki}(p^2) -
      \hat{\Gamma}^2_{ki}(p^2) \hat{\Gamma}_{jj}(p^2) -
      \hat{\Gamma}^2_{ij}(p^2) \hat{\Gamma}_{kk}(p^2)
     }{\hat{\Gamma}_{jj}(p^2) \hat{\Gamma}_{kk}(p^2) - 
       \hat{\Gamma}^2_{jk}(p^2)
      } ,
\label{eq:sigmaeff}
\end{align}
where the $\hat{\Gamma}_{ij}(p^2)$ are the elements of the $3 \times 3$
matrix $\hat{\Gamma}_{hHA}(p^2)$ as specified in
\refeq{eq:invprophiggs}.
The complex pole is decomposed as 
\begin{equation}
{\cal M}^2 = M^2 - i M \Ga ,
\end{equation}
where $M$ is the mass of the particle and $\Ga$ its width,
We define the loop-corrected mass eigenvalues according to
\begin{equation}
M_{h_1} \leq M_{h_2} \leq M_{h_3} .
\label{eq:mh123}
\end{equation}

In our determination of the Higgs-boson masses we take into account all
imaginary parts of the Higgs-boson self-energies 
(besides the term with imaginary parts 
appearing explicitly in \refeq{eq:massmaster}, there are also products
of imaginary parts in $\re \ser{ii}^{\rm eff}(M_i^2)$).
The effects of the imaginary parts of the Higgs-boson self-energies on
Higgs phenomenology can be 
especially relevant if the masses are close to each other. 
This has been analyzed in \citere{imagSE1} taking into account the
mixing between the two heavy neutral Higgs bosons, where the complex
mass matrix has been diagonalized using a complex mixing angle,
resulting in a non-unitary mixing matrix.
The effects of imaginary parts of the Higgs-boson self-energies on
physical processes with s-channel resonating Higgs bosons
are discussed in \citeres{imagSE1,imagSE2,imagSE3}. In
\citere{imagSE1} only the one-loop corrections from the
$t/\Stop$~sector have been taken into account for the $H$--$A$~mixing,
analyzing the effects on resonant Higgs production at a photon collider.
In \citere{imagSE2} (using the code {\tt CPSuperH}~\cite{cpsh}) 
the full one-loop imaginary parts of the
self-energies have been evaluated for the mixing of the three neutral
MSSM Higgs bosons. The effects have been analyzed for resonant Higgs
production at the LHC, the ILC and a photon collider (however, the
corresponding effects on the Higgs-boson masses have been neglected). 
In \citere{imagSE3} the $\Stop/\Sbot$ one-loop contributions (neglecting the
$t/b$ corrections) on the $H$--$A$ mixing for resonant Higgs
production at a muon collider have been discussed.
Our calculation~\cite{mhcMSSMlong,feynhiggs2.5} incorporates for the first
time the 
complete effects arising from the imaginary parts of the one-loop
self-energies in the neutral Higgs-boson propagator
matrix, including their effects on the Higgs masses and the Higgs
couplings in a consistent way.


\section{Numerical example}

In order to study the impact of the imaginary parts of the Higgs-boson
self-energies, it is useful to compare the full result with the 
``$\im\Si = 0$'' approximation, which is defined by performing 
the replacement
\begin{equation}
\mbox{$\im\Si = 0$ approximation: } \quad
\begin{array}{lcl}
\Si(p^2) & \to & \re \Si(p^2) 
\end{array}
\label{eq:ImSi0approx}
\end{equation}
for all Higgs-boson self-energies in \refeq{eq:Mn}. 
The numerical example has been obtained for the following set of
parameters:
\BEA
&& \msusy = 500 \gev, \; |\At| = \Ab = \Atau = 1000 \gev, 
   \; \non \\
&& \mu = 1000 \gev, \; M_2 = 500 \gev, \; M_1 = 250 \gev, \;
\mgl = 500 \gev, \non \\
&& \mudim = \mt = 171.4 \gev~\mbox{\cite{mt1714}}.
\label{parameters}
\EEA

\begin{figure}[htb!]
\centerline{\includegraphics{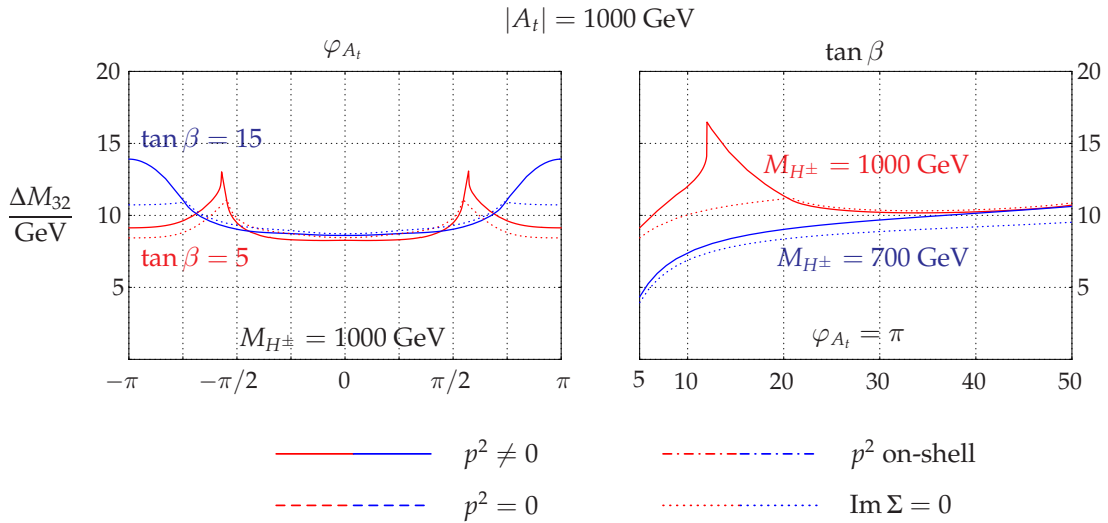}}
\caption{
The mass difference $\De M_{32} := \MHd - \MHz$ is shown for 
$\tb = 5, 15$ and $\MHp = 1000 \gev$ 
as a function of $\phiat$ (left) and for $\phiat = \pi$, 
$\MHp = 700, 1000 \gev$ as a function of $\tb$ (right). 
The solid line shows the full result, the dotted line the ``$\im\Si = 0$''
approximation. The other two lines correspond to other Higgs-boson
self-energy approximations, see \citere{mhcMSSMlong} for details.
}
\label{fig:imSi0}
\end{figure}

In \reffi{fig:imSi0} we show an example of
the effects of the imaginary parts of the Higgs-boson
self-energies, i.e.\ the comparison of the full result with the 
``$\im\Si = 0$'' approximation as defined in \refeq{eq:ImSi0approx}.
In the left plot we show $\De M_{32} := \MHd - \MHz$ as a function
of $\phiat$ for $\tb = 5, 15$ and $\mHp = 1000 \gev$. 
In the right plot we
display $\De M_{32}$ as a function of $\tb$ for $\mHp = 700, 1000 \gev$ and
$\phiat = \pi$. Here $\phiat$ denotes the angle of the complex valued
trilinear coupling $\At$.
The other parameters are given in \refeq{parameters}. 
As one can see from the plot, the difference between the full result
and the approximation with neglected imaginary parts is often 
$\sim 1 \gev$ and can become as large as about $5 \gev$.


\section{Acknowledgments}

We thank K.~Williams for numerous checks and helpful discussions.
Work supported in part by the European Community's Marie-Curie Research
Training Network under contract MRTN-CT-2006-035505
`Tools and Precision Calculations for Physics Discoveries at Colliders'




\begin{footnotesize}


\end{footnotesize}


\begin{thebibliography}{99}
\bibitem{url} Slides: \\ 
\verb$ilcagenda.linearcollider.org/contributionDisplay.py?contribId=158&sessionId=71&confId=1296$

\bibitem{mssm} H.~Nilles, 
               {\em Phys.\ Rept.} {\bf 110} (1984) 1;
               H.~Haber and G.~Kane, 
               {\em Phys.\ Rept.} {\bf 117} (1985) 75; 
               R.~Barbieri, 
               {\em Riv.\ Nuovo Cim.} {\bf 11} (1988) 1. 

\bibitem{LEPHiggsSM} [LEP Higgs working group],
                     {\em Phys. Lett.} {\bf B 565} (2003) 61
                     [arXiv:hep-ex/0306033].

\bibitem{LEPHiggsMSSM} [LEP Higgs working group],
                       {\em Eur.\ Phys.\ J.} {\bf C 47} (2006) 547
                       [arXiv:hep-ex/0602042].

\bibitem{ERZ} J.~Ellis, G.~Ridolfi and F.~Zwirner,
              {\em Phys.\ Lett.} {\bf B 257} (1991) 83;
              Y.~Okada, M.~Yamaguchi and T.~Yanagida,
              {\em Prog.\ Theor.\ Phys. } {\bf 85} (1991) 1;
              H.~Haber and R.~Hempfling,
              {\em Phys.\ Rev.\ Lett.}  {\bf 66} (1991) 1815.

\bibitem{mhiggsf1lA} A.~Brignole,
                     {\em Phys. Lett.}\ {\bf B 281} (1992) 284.

\bibitem{mhiggsf1lB} P.~Chankowski, S.~Pokorski and J.~Rosiek,
                     {\em Phys. Lett.} {\bf B 286} (1992) 307;
                     {\em Nucl. Phys.} {\bf B 423} (1994) 437
                     [arXiv:hep-ph/9303309].

\bibitem{mhiggsf1lC} A.~Dabelstein,
                     {\em Nucl. Phys.} {\bf B 456} (1995) 25
                     [arXiv:hep-ph/9503443];
                     {\em Z. Phys.} {\bf C 67} (1995) 495
                     [arXiv:hep-ph/9409375].

\bibitem{mhiggsAEC} G.~Degrassi, S.~Heinemeyer, W.~Hollik,
                    P.~Slavich and G.~Weiglein, 
                    {\em Eur. Phys. J.} {\bf C 28} (2003) 133
                    [arXiv:hep-ph/0212020].

\bibitem{mhiggsEP5} S.~Martin, 
                    {\em Phys. Rev.} {\bf D 71} (2005) 016012
                    [arXiv:hep-ph/0405022];
                    {\em Phys. Rev.} {\bf D 71} (2005) 116004
                    [arXiv:hep-ph/0502168].

\bibitem{mhiggs3L} S.~Martin, 
                   {\em Phys.\ Rev.} {\bf D 75} (2007) 055005
                   [arXiv:hep-ph/0701051].

\bibitem{PomssmRep} S.~Heinemeyer, W.~Hollik and G.~Weiglein,
                    {\em Phys.\ Rept.} {\bf 425} (2006) 265
                    [arXiv:hep-ph/0412214].

\bibitem{mhiggsWN} B.~Allanach, A.~Djouadi, J.~Kneur, W.~Porod and P.~Slavich,
                   {\em JHEP} {\bf 0409} (2004) 044
                   [arXiv:hep-ph/0406166].

\bibitem{mhiggslong} S.~Heinemeyer, W.~Hollik and G.~Weiglein,
                     {\em Eur. Phys. J.} {\bf C 9} (1999) 343
                     [arXiv:hep-ph/9812472].

\bibitem{feynhiggs} S.~Heinemeyer, W.~Hollik and G.~Weiglein,
                    {\em Comput. Phys. Commun.} {\bf 124} (2000) 76
                    [arXiv:hep-ph/9812320]; 
                    see {\tt www.feynhiggs.de} .

\bibitem{mhcMSSMlong} M.~Frank, T.~Hahn, S.~Heinemeyer, W.~Hollik, 
                      R.~Rzehak and G.~Weiglein,
                      {\em JHEP} {\bf 02} (2007) 047
                      [arXiv:hep-ph/0611326].

\bibitem{mhiggsletter} S.~Heinemeyer, W.~Hollik and G.~Weiglein, 
                       {\em Phys. Rev.} {\bf D 58} (1998) 091701
                       [arXiv:hep-ph/9803277]; 
                       {\em Phys. Lett.} {\bf B 440} (1998) 296 
                       [arXiv:hep-ph/9807423].

\bibitem{mhiggsEP4b} G.~Degrassi, A.~Dedes and P.~Slavich,
                    {\em Nucl. Phys.} {\bf B 672} (2003) 144
                    [arXiv:hep-ph/0305127].

\bibitem{mhcMSSM2L} S.~Heinemeyer, W.~Hollik, H.~Rzehak and G.~Weiglein,
                    to appear in {\em Phys. Lett.} {\bf B}, 
                    arXiv:0705.0746 [hep-ph].

\bibitem{imagSE1} S.~Choi, J.~Kalinowski, Y.~Liao and P.~Zerwas,
                  {\em Eur.\ Phys.\ J.} {\bf C 40} (2005) 555 
                  [arXiv:hep-ph/0407347].

\bibitem{imagSE2} J.~Ellis, J.~Lee and A.~Pilaftsis,
                  {\em Phys.\ Rev.} {\bf D 70} (2004) 075010
                  [arXiv:hep-ph/0404167];
                  {\em Nucl.\ Phys.} {\bf B 718} (2005) 247
                  [arXiv:hep-ph/0411379];
                  {\em Phys.\ Rev.} {\bf D 72} (2005) 095006
                  [arXiv:hep-ph/0507046].

\bibitem{imagSE3} J.~Bernabeu, D.~Binosi and J.~Papavassiliou,
                  {\em JHEP} {\bf 0609} (2006) 023
                  [arXiv:hep-ph/0604046].

\bibitem{cpsh} J.~Lee, A.~Pilaftsis et al.,
               {\em Comput. Phys. Commun.} {\bf 156} (2004) 283
               [arXiv:hep-ph/0307377].

\bibitem{feynhiggs2.5} T.~Hahn, S.~Heinemeyer, W.~Hollik, H.~Rzehak, 
                       G.~Weiglein and K.~Williams,
                       arXiv:hep-ph/0611373.

\bibitem{mt1714} E.~Brubaker et al. [Tevatron Electroweak Working Group],
                 arXiv:hep-ex/0608032, 
                 see:\\ {\tt tevewwg.fnal.gov/top/}~.


\end{thebibliography}
\end{document}